\documentstyle[11pt,epsfig]{article}
[1999/03/29 PSNFSS v.7.2
Times font as default roman
: S Rahtz]

\begin{document}
\renewcommand{\thefootnote}{\fnsymbol{footnote}}
\sloppy
\newcommand{\rp}{\right)}
\newcommand{\lp}{\left(}
\newcommand \be  {\begin{equation}}
\newcommand \bea {\begin{eqnarray}}
\newcommand \ee  {\end{equation}}
\newcommand \eea {\end{eqnarray}}

\title{Self-Consistent Theory of Rupture by Progressive Diffuse Damage}
\thispagestyle{empty}

\author{S. Gluzman$^1$ and D. Sornette$^{1,2,3}$\\
$^1$ Laboratoire de Physique de la Mati\`{e}re Condens\'{e}e\\ CNRS UMR6622 and
Universit\'{e} de Nice-Sophia Antipolis\\ B.P. 71, Parc
Valrose, 06108 Nice Cedex 2, France
$^2$ Institute of Geophysics and
Planetary Physics\\ University of California, Los Angeles, California 90095\\
$^3$ Department of Earth and Space Science\\
University of California, Los Angeles, California 90095\\
e-mails: gluz@idirect.com and sornette@unice.fr \\}

\maketitle

\begin{abstract}
We analyze a self-consistent theory of crack growth controlled by a
cumulative damage variable $d(t)$
dependent on stress history. As a function
of the damage exponent $m$, which controls the rate of damage $dd/dt \propto \sigma^m$
as a function of local stress $\sigma$, we find two regimes.
For $0 < m < 2$, the model predicts a finite-time singularity. This retrieves
previous results by Zobnin for $m=1$ and by Bradley and Wu for $0 < m < 2$.
To improve on this self-consistent
theory which neglects the dependence of stress on damage, 
we apply the functional renormalization method of Yukalov and Gluzman 
and find that divergences are replaced
by singularities with exponents in agreement with those found in acoustic emission experiments.
For $m \geq 2$, the 
rupture dynamics is not defined without the introduction of a regularizing scheme.
We investigate three regularization schemes involving respectively a saturation of damage,
a minimum distance of approach to the crack tip and a fixed stress maximum. 
In the first and third schemes, the finite-time singularity is replaced by a crack
dynamics defined for all times but which is
controlled by either the existence of a microscopic scale at which the stress is regularized or
by the maximum sustainable stress.
In the second scheme, a finite-time singularity is again found.
In the first two schemes within this regime $m \geq 2$, the theory has no continuous limit.

\end{abstract}

\pagenumbering{arabic}

\section{Introduction}

The fracture of materials is a catastrophic phenomenon of considerable 
technological and scientific importance. Despite the large amount of
experimental data and the considerable effort that has been undertaken by
material scientists \cite{inge}, many questions about fracture remain standing.
There is no comprehensive understanding of rupture
phenomena but only a partial classification in
restricted and relatively simple situations. This lack of fundamental
understanding is reflected in the absence of reliable prediction
methods for rupture based on a suitable monitoring of the stressed  system. 

Some progresses have been obtained in recent years in the Physics community. Based
on analogies with phase transitions, several groups \cite{Anifrani}-\cite{criticalrecent} have proposed
that, in heterogeneous materials with disorder such as fiber composites,  rocks, concrete 
under compression and materials with large distributed residual stresses, 
rupture is a genuine critical point, {\it i.e.}, the culmination of a 
self-organization of diffuse damage and micro-cracking characterized by power law 
signatures. Experiments \cite{Anifrani,Lamai,Ciliberto,Anifrani2,criticalrecent},
numerical simulations \cite{Vanneste,Arbati,Andersen1,Andersen2} and theory
\cite{Andersen1} confirm this concept.

As a signature of criticality, acoutic emissions radiated during loading
exhibit an acceleration of their rate close to rupture \cite{Anifrani,Ciliberto,criticalrecent}. Specifically,
under a constant stress rate,
the cumulative acoustic energy $E(t)$ released up to time $t$ can be expressed as
\be
E(t) = E_0 - B (t_c-t)^{\alpha}~,   \label{one}
\ee
with $B>0$ and $0 < \alpha < 1$. Expression (\ref{one}) corresponds to a rate $dE/dt$ of acoustic energy release
diverging at the critical rupture time $t_c$. This behavior (\ref{one}) has been at the basis
of previous claims that rupture is a critical phenomenon. In addition, this power law (\ref{one})
as well as extensions with log-periodic corrections have been suggested to be useful for prediction
\cite{Voight,Anifrani,Sammis,Anifrani2,criticalrecent}.

Our purpose here is to present,  extend and analyze a simple self-consistent model of damage that
predicts a behavior similar to (\ref{one}). We explore its different regimes
and then improve on its  ``mean-field'' version which predicts
an unrealistic finite-time singularity. In this goal, we propose to use the general functional
renormalization approach developed by Yukalov and Gluzman to cure this anomaly. 
We show how this technique allows us to change an unrealistic singularity into the observed
behavior (\ref{one}) with a reasonable exponent $z=1/2$, without introduction of any 
extra parameters in the theory.

\section{Cumulative damage model}

Initially introduced as a global
``mean field'' (uniform) description of the global deterioration of the system at the 
macroscopic scale \cite{Chaboche}, 
the concept of ``damage'' has been extended at the mesoscopic scale to describe 
the heterogeneity and spatial variability of damage in different locations within the material
\cite{yang,Vanneste,mechaanalog,benz}. We use the formulation of Zobnin 
\cite{Zobnin} and Rabotnov \cite{Rabotnov} to show how
it leads naturally to a finite-time singularity. We first recall briefly the integral formulation
of Rabotnov \cite{Rabotnov} (pages 166-170) and then transform it in differential form to 
exhibit the fundamentally nonlinear geometrical origin of the singularity. 

A material is subjected to a stress $\sigma_0$ at large scale and each point $r$ within it carries
a damage variable $d(r,t)$. When $d$ reaches the threshold $d^*$ as some location, this local domain
is no more able to sustain stress and a microcrack appears, leading to a redistribution of the stress
field around it according to the laws of elasticity.
 The local damage $d(r,t)$ at point $r$ at time $t$ is supposed to evolve 
 in time according to
\be
{d (d) \over dt} = [\sigma(r,t)]^m~,  \label{two}
\ee
where $\sigma(r,t)$ is the local stress field at point $r$ at time $t$ and $m$ is a damage exponent
which can span values from $0$ to close to $\infty$ depending upon the material. In 
the discrete 2D models of Refs.\cite{Vanneste,mechaanalog},
it was shown that rupture reduces to the percolation model in the limit $m \to 0$. In the other 
limit, $m \to \infty$, rupture occurs through a one crack mechanism.

Following Rabotnov \cite{Rabotnov}, we assume that a major crack dominates the rupture process.
If only one crack is present within the system, the stress $\sigma(r,t)$ is easily
calculated. Considering only the possibility of a linear straight crack of half-length
$a(t)$ advancing within the material at velocity $da/dt$ (see 
\cite{roux} for generalizations to self-affine crack geometries), it is enough to calculate the stress field
on the points ahead of the crack to fully characterize the rupture dynamics. 
For a planar elastic material subjected to a uniformly distributed antiplane stress at 
infinity with a crack lying on the $y$-axis between $-a(t)$ and $a(t)$, 
the stress field at point $y$ 
on the $y$-axis beyond the crack tip is
\be
\sigma(z,t) = {2 \sigma_0 \over 3} {z \over \sqrt{z^2 - [a(t)]^2}}~.   \label{three}
\ee
The mean-field approximation made in this first version of the model consists in assuming
 that the stress field
is not modified by the non-vanishing and non-uniform damage field. This means that the elastic
coefficients are taken constant and independent of the progressive damage, except of course
when the damage reaches its rupture threshold $d^*$.

The law describing the growth of the crack, i.e., the dynamics $a(t)$, is obtained from the
self-consistent condition that the time it takes from a point at $y$, at the distance $y-a(\tau)$
from the crack tip at time $\tau$, for its damage to reach the rupture threshold $d^*$ is
exactly equal to the time taken for the crack to grow from size $a(\tau)$ to the size $a(t)=y$
so that its tip reaches the point $y$ exactly when it ruptures. This is
illustrated in figure 1.
Mathematically, this self-consistent condition is that
the integral of (\ref{two}) from time $0$ at which the pre-existing damage was $0$ till 
time $t$ at which the crack tip passes through $y$ is such that $d$ reaches exactly the threshold $d^*$
at the time $t$. Two conditions must thus be verified simultaneously: 
\begin{enumerate}
\item $a(t) = y$ (the crack tip reaches point  $y$) and
\item $d(y,t)=d^*$ (the damage at $y$ reaches the rupture threshold). 
\end{enumerate}

\section{The linear damage law: $m=1$}

We first consider the linear damage law $m=1$ corresponding to the initial
formulation of Zobnin \cite{Zobnin}. This case has also been investigated and solved
in \cite{Bradley1} in the context of crack growth due to electromigration
rather than mechanical stress (the current 
plays the role of the stress and, in the antiplane case studied here,
the two problems are formally identical).
This model is 
particularly interesting since it allows both for 
an exact solution and an exact renormalization in the functional
renormalization scheme \cite{YG}. It also 
 provides a benchmark for approximate solutions in the general case $0<m<2$
 as we discuss below.
 
We now proceed to give the equation for the crack dynamics and its solution.
By integration of (\ref{two}), the two self-consistent conditions expressed 
for the case $m=1$ lead to
\be
\int_0^t d\tau~ {2 \sigma_0 \over 3} {a(t) \over \sqrt{[a(t)]^2 - [a(\tau)]^2}} = d^*~,   \label{jbba}
\ee
where the loading stress $\sigma_0$ can depend on time.
The solution of this integral equation provides the time evolution $a(t)$ of the macro-crack. To get it
explicitely, we set 
\be
z = [a(t)]^2~~~~~~{\rm and}~~~~~~ \zeta=[a(\tau)]^2~.   \label{nfvnkzka}
\ee
Changing the variable of integration from $\tau$ to $\zeta$ gives
\be
\int_{z_0}^{z} d\zeta~ \sigma_0 {(d\tau/d\zeta) \over \sqrt{z-\zeta}} = {3 \over 2} {d^* \over
\sqrt{z}}~.  \label{cnnkz}
\ee
This equation (\ref{cnnkz}) is an Abel equation with index $-1/2$, 
involving a fractional integral operator \cite{Hilfer}. Defining the Abel operator $I^*_{\alpha}$
acting on the function $f(t)$ as
\be
I^*_{\alpha}\{f\} = \int_0^t {(t-s)^\alpha \over \Gamma(1+\alpha)}~ f(s) ds~,
\ee
the product of two such Abel operators is
$$
I^*_{\alpha}~I^*_{\beta}\{f\} = I^*_{\alpha}\{I^*_{\beta}\{f\}\} =
\int_0^t d\tau~{(t-\tau)^\alpha \over \Gamma(1+\alpha)}~
  \int_0^{\tau} {(\tau-s)^\beta \over \Gamma(1+\beta)}~ f(s) ds~ 
$$
\be
= \int_0^{t} ds~ f(s) ~ {1 \over \Gamma(1+\alpha)\Gamma(1+\beta)} \int_s^t d\tau~
(t-\tau)^\alpha~(\tau-s)^\beta~ = \int_0^t ds ~f(s) {(t-s)^{\alpha+\beta +1} \over \Gamma(2+\alpha+\beta)}~.
\ee
This shows that 
\be
I^*_{\alpha}~I^*_{\beta}\{f\} = I^*_{\alpha+\beta+1}\{f\}~.
\ee
We thus see that $I^*_{\alpha}~I^*_{-1-\alpha}\{f\} = I^*_{0}$, which is nothing but the 
integral operator. 
The inverse of the Abel operator $I^*_{\alpha}$ is thus ${d \over dt}\left(I^*_{-1-\alpha}\right)$.
Applying this result to (\ref{cnnkz}), we find
\be
{d \tau \over dz} = {3 \over 2} {d^* \over \pi \sigma_0}~ {d \over dz} 
\int_{z_0}^{z}   {d\zeta \over \sqrt{\zeta(z-\zeta)}} ~.  \label{faaaa}
\ee
Calculating the integral in the r.h.s. of (\ref{faaaa}),
performing the derivative and inverting to get $dz/d\tau$, we get
\be
{dz \over d\tau} = {2 \pi \sigma_0 \over 3 d^*}~ {z \sqrt{z-z_0} \over \sqrt{z_0}}~.
\ee
Replacing $z$ by $[a(t)]^2$ leads to the differential equation for the crack half-length $a(t)$
\be
{da \over dt} = {\pi \sigma_0 \over 3 d^*}~ a ~\sqrt{\left({a \over a_0}\right)^2 -1}~, \label{mnakks}
\ee
which is exactly 
equivalent to the self-consistent integral equation (\ref{jbba}). It is remarkable that
the local growth equation (\ref{mnakks}) embodies exactly the same physics as the long-term
memory integral (\ref{jbba}).

For simplicity, let us take the loading stress $\sigma_0$ constant. This situation is generic of
experiments measuring the lifetime of structures under a constant load.
At sufficiently long times for which $a(t) >> a_0$, expression (\ref{mnakks}) reduces to 
\be
{da \over dt} \approx {\pi \sigma_0 \over 3 a_0 d^*}~ a^2 ~. \label{mnakkaaas}
\ee
Equation (\ref{mnakkaaas}) is characteristic of a solution going to infinity in finite time. 
Indeed, we can write (\ref{mnakkaaas}) as $da/dt \propto r a$, with a growth rate $ r \propto a$.
The generic consequence of a power law acceleration in the growth rate $ r \propto a^{\delta}$
with $\delta >0$ is the
appearance of a singularity in finite time:
\be \label{pow}
a(t) \propto (t_c - t)^{-\beta}, ~~{\rm with}~\beta={1 \over \delta}~~ \mbox{ and $t$
close to $t_c$}.
\ee
Equation (\ref{mnakkaaas}) is said to have a ``spontaneous'' or ``movable''
singularity at the critical time $t_c$ \cite{benderorszag}, the critical time
$t_c$ being determined by the constant of integration, {\it i.e.}, the initial
condition $a(t=0)=a_0$.  Note the intriguing fact that the $(t_c-t)^{-1}$ singularity
appears as the solution of a linear mechanical problem. The source of the quadratic nonlinearity is
the non-local geometrical condition that the
delayed action of the stress field on the cumulative damage should coincide exactly with 
the passage of the crack tip. The nonlinear finite-time singularity has thus fundamentally
a non-local geometrical origin, or alternatively can be seen to result from a long-term
memory effect.

The exact solution of (\ref{mnakks}) is easily obtained by integration:
\be
a(t) = {a_0 \over \cos \left({\pi \sigma_0 \over 3 d^*}~t\right)}~.  \label{njjakak}
\ee
This retrieves the solution obtained by Zobnin \cite{Zobnin} and Rabotnov \cite{Rabotnov}.
We verify directly that the singularity occurs when the
cosine goes to zero, i.e., when the argument reaches $\pi/2$, i.e., for $t_c= 3d^*/2\sigma_0$.
Since the cosine vanishes linearly with time,
this recovers the asymptotics (\ref{pow}) with the exponent $\beta=-1$, as predicted by the 
asymptotic equation (\ref{mnakkaaas}).

\section{The nonlinear damage law with finite-time singularity: $0 < m < 2$}

\subsection{Derivation of the differential equation for the crack dynamics}

The case where $0 < m < 2$ can be similarly treated and our results here 
extend those of Zobnin \cite{Zobnin} and Rabotnov \cite{Rabotnov}. Our results retrieve
those found in \cite{Bradley2}, obtained in the context of crack growth due to electromigration.
For completeness and coherence in notation, we briefly present the method
and the results which are more focused on the finite-time singularity.

For simplicity, we impose $\sigma_0$ constant.
Integrating (\ref{two}) and applying the self-consistent conditions 1-2 leads to
\be
\int_0^t d\tau~ \left({2 \sigma_0 \over 3}\right)^{m}
 {[a(t)]^m \over \left([a(t)]^2 - [a(\tau)]^2\right)^{m \over 2}} = d^*~.   \label{jbbaaa}
\ee
We set again the change of variables (\ref{nfvnkzka}) and
changing the variable of integration from $\tau$ to $\zeta$ gives
\be
\int_{z_0}^{z} d\zeta~ {(d\tau/d\zeta) \over \left[z-\zeta\right]^{m \over 2}} = 
\left({3 \over 2 \sigma_0}\right)^m {d^* \over z^{m \over 2}}~.  \label{cnnkfdaz}
\ee
This equation (\ref{cnnkz}) is again an Abel equation with index $-m/2$ if $0 < m < 2$.

In order to transform it into differential form, we could use the formalism of Abel operators.
We choose a more transparent and direct approach which is closely related. First, we multiply
both sides of (\ref{cnnkfdaz}) by $1/(y-z)^{1-{m \over 2}}$ and integrate over $z$ from $z_0$ to $y$:
\be
\int_{z_0}^{y} dz \left(\int_{z_0}^{z} d\zeta~ {(d\tau/d\zeta) \over 
\left(z-\zeta\right)^{m \over 2} (y-z)^{1-{m \over 2}} }\right) =
\left({3 \over 2 \sigma_0}\right)^m d^* 
\int_{z_0}^{y} {dz \over  z^{m \over 2} (y-z)^{1-{m \over 2}}}~.
\ee
Changing the order of integration in the l.h.s. leads to 
\be
\int_{z_0}^{z} d\zeta {d\tau \over d\zeta} \left[
\int_{\zeta}^{y} {dz  \over \left(z-\zeta\right)^{m \over 2} (y-z)^{1-{m \over 2}}}\right] =
\left({3 \over 2 \sigma_0}\right)^m d^* 
\int_{z_0}^{y} {dz \over  z^{m \over 2} (y-z)^{1-{m \over 2}}}~,  \label{cbvvc}
\ee
where we have used the equality of the triangle 
$\int_{z_0}^{y} dz  \int_{z_0}^{z} d\zeta = \int_{z_0}^{z} d\zeta \int_{\zeta}^{y} dz$.

The integral in the square bracket in the l.h.s. of (\ref{cbvvc}) can be expressed through the Euler beta-function
$B(a,b)$:
\be
\int_{\zeta}^{y} {dz  \over \left(z-\zeta\right)^{m \over 2} (y-z)^{1-{m \over 2}}} =
B\left(1-{m \over 2}, {m \over 2}\right) = {\Gamma\left(1-{m \over 2}\right)~\Gamma\left({m \over 2}\right)
\over \Gamma\left(1-{m \over 2} + {m \over 2}\right)} = {\pi \over \sin\left(m {\pi \over 2}\right)}~. \label{bcvvvc}
\ee
We thus obtain
\be
{\pi \over \sin\left(m {\pi \over 2}\right)} \int_{z_0}^{z} d\zeta {d\tau \over d\zeta} =
\left({3 \over 2 \sigma_0}\right)^m d^* 
\int_{z_0}^{y} {dz \over  z^{m \over 2} (y-z)^{1-{m \over 2}}}~.  \label{cbaafdvvc}
\ee
After differentiation with respect to $z$, we get
\be
{d \tau \over dz} = \left({3 \over 2 \sigma_0}\right)^m d^*~{\sin\left(m {\pi \over 2}\right) \over \pi}~ {d \over dz} 
\int_{z_0}^{z}   {d\zeta \over \zeta^{m \over 2} (z-\zeta)^{1-{m \over 2}}} ~.  \label{faadafaaaaa}
\ee

\subsection{Asymptotic solution close to the finite-time singularity}

The solution of (\ref{faadafaaaaa}) can be obtained for large crack sizes $a(t)$, i.e., large $z$. 
In this goal, we replace the term $(z-\zeta)^{1-{m \over 2}}$ in the integral in the r.h.s. of (\ref{faadafaaaaa})
by $z^{1-{m \over 2}}$, neglecting $\zeta$ compared to $z$. Intuitively, this is justified 
over the whole domain of integration because the contribution from the domain where $\zeta$ is not negligible 
compared to $z$ is finite, since the power $1-{m \over 2}$ is less than one, corresponding to 
an integrable singularity. 

With this approximation, the integral can be performed, the derivative taken and after inverting, we get
\be
\left({3 \over 2 \sigma_0}\right)^m d^*~{\sin\left(m {\pi \over 2}\right) \over \pi}~ {d z \over d\tau} = 
 z_0^{-1+{m \over 2}} ~z^{2 -{m \over 2}}~.  \label{faaqqqaa}
\ee
Using $a=\sqrt{z}$ as defined in (\ref{nfvnkzka}), we obtain
\be
a(t) = {a_0 \over \left(1 - {t \over t_c}\right)^{\beta}}~, \label{hgkzz}
\ee
where 
\be
t_c = \left({3 \over 2\sigma_0}\right)^m d^*~{2\sin\left(m {\pi \over 2}\right) \over \pi (2-m)}~, \label{bvcbnxx}
\ee
and
\be
\beta = {1 \over 2 - m}~.
\ee

Note that the exact asymptotics (\ref{pow}) of the case $m=1$ previously solved exactly is recovered, 
with the correct exponent $z(m=1)=1$ and a rather good approximation of the critical $t_c$: while
the exact value is $t_c=3 d^*/3\sigma_0$, expression (\ref{bvcbnxx}) predicts ${2 \over \pi} t_c$, i.e.,
$36\%$ lower. The critical time $t_c$ as a function of $m$  is 
smooth with no accident or divergence over the whole interval. In particular, the estimated critical time
for the limit $m \to 2^-$ is equal to $9d^*/4 \sigma_0^2$.

In contrast, the exponent $z$ increases from $z(m \to 0^+)=1/2$ to $+\infty$ as the damage
exponent $m$ varies from $0$ to $2$. The limit $z(m \to 0^+)=1/2$ can be rationalized as follows.
This limit $m \to 0^+$ corresponds to the situation where damage becomes independent of stress. As
a consequence, reintroducing some heterogeneity for instance on the pre-existing damage, rupture
is then equivalent to percolation, as the parts of the system that break as a function of time
 are determined by the damage accumulating at the same rate for all point but with different random
initial values. In
mean field percolation \cite{perco} obtained through the consideration of one-dimensional 
percolating paths consistent with the present one-crack geometry,
the elastic energy under constant load diverges a
$(t_c-t)^{-1}$ where $1$ is the mean field value of the exponent $t$ for conductivity (which
is the same as elasticity in the scalar mode III version of mechanical
deformations used here). Since the elastic energy is proportional to the square of the crack
length, we get the prediction $a(t) \sim (t_c-t)^{-1/2}$. This reasoning holds if
the exponent is a smooth function of disorder and geometry (the present studied here
is a the zero-disorder limit).

The divergence of $z$ at $m=2$ signals a change of regime that
we study in the next section.

\section{The nonlinear damage law with $m \geq 2$}

For $m \geq 2$, the integrals in the equations (\ref{jbbaaa}) and (\ref{cnnkfdaz}) diverge
at $a(\tau)=a(t)$, since the
negative power with exponent $m/2$ is no more integrable. Technically, the main difference between the
cases $m<2$ and $m \geq 2$ is that we need to regularize the infinity in the expression (\ref{bcvvvc}) by
introducing some sort of dimensionless cut-off. The important physical message is that the regime where $m \geq 2$
is controlled by a novel physical parameter, which we identify as a length scale associated with the
damage law. In other words, the physics of the rupture is inherently controlled by the choice of the
cut-off, i.e., by the existence of a microscopic length scale. We could summarize the situation by
saying that there is no continuous limit to the theory for $m \geq 2$. This is similar to previous observations
obtained in a dynamical theory of rupture front propagation \cite{Langer}.
We now present two ways for regularizing the divergence and thus for obtaining a meaningful theory of rupture.

\subsection{Regularization by damage saturation at a microscopic scale}

Before describing the physical content of the regularization we propose, we need to express the 
problem in a more manageable mathematical form. Since the culprit for the divergence is the integral
(\ref{bcvvvc}) and the divergence occurs for $\zeta \to z^-$,  we introduce the variable
\be
Z= {z-\zeta \over y-\zeta}~,  \label{nbfblva}
\ee
and rewrite (\ref{bcvvvc}) as
\be
\int_{\zeta}^{y} {dz  \over \left(z-\zeta\right)^{m \over 2} (y-z)^{1-{m \over 2}}} =
\int_0^1 dZ ~Z^{-{m \over 2}}~ (1-Z)^{{m \over 2}-1}~,  \label{bvbbhsk}
\ee
which makes apparent that the divergence is due to $Z^{-{m \over 2}}$ at the lower bound $0$.
It is thus natural to regularize by introducing a dimensionless cut-off $\epsilon>0$ and replace
(\ref{bvbbhsk}) by
\be 
\int_{\epsilon}^1 dZ ~Z^{-{m \over 2}}~ (1-Z)^{{m \over 2}-1} \equiv b(m,\epsilon)~.  \label{bvabbhsk}
\ee
The function $b(m,\epsilon)$ is such that
\be
{\rm lim}_{\epsilon \to 0^+}~b(m,\epsilon) = B(1-{m \over 2}, {m \over 2})~,~~~~~~{\rm for}~~0 < m < 2~,
\ee
where the beta function $B(1-{m \over 2}, {m \over 2})$ has been defined in (\ref{bcvvvc}).

In constrast, we have
\be
b(m,\epsilon) \sim {1 \over \epsilon^{m - 2 \over 2}}~,~~~~~{\rm for}~~m>2~,
\ee
and 
\be
b(m,\epsilon) \sim \ln {1 \over \epsilon}~,~~~~~{\rm for}~~m=2~,
\ee
showing that the divergence of the integral (\ref{bvbbhsk}) is now encapsulated in the 
dependence of the factor $b(m,\epsilon)$ on $\epsilon$. 
This regularization scheme thus relies on the
existence of the definite integral (\ref{bvabbhsk}), by analogy to the case $m<2$.

Using the regularization (\ref{bvabbhsk}), we obtain
\be
{d \tau \over dz} = \left({3 \over 2\sigma_0}\right)^m ~{d^* \over b(m,\epsilon) }~ {d \over dz} 
\int_{z_0}^{z}   {d\zeta \over \zeta^{m \over 2} (y-\zeta)^{1-{m \over 2}}} ~,  \label{faaaraaa}
\ee
which extends (\ref{faaaa}) to the regime $m\geq 2$. Its formal solution obtained in implicit form is
\be
t = \left({3 \over 2\sigma_0}\right)^m~{d^* \over b(m,\epsilon)}~
\int_{z_0}^z dy {d \over dy} \int_{z_0}^{y}   {d\zeta \over \zeta^{m \over 2} (y-\zeta)^{1-{m \over 2}}}~. \label{nvbnms}
\ee

This regularization scheme allows to obtain exact solutions for integer $m$'s. 
We examine the solutions for $m=2, 3, 4$ and $5$ and then the general case.
 For m=2, we have the expression for all times given by
 \be
z_2(t) = z_0 e^{t/t_{\epsilon,2}}~,
\ee
where 
\be
t_{\epsilon,2} = {9 d^* \over 4 \sigma_0^2}~{1 \over \ln {1 \over \epsilon}}~.
\ee

For $m=3$, $z_3(t)$ is the solution to
\be
\sqrt{(z/z_0)-1} - \tan^{-1} (\sqrt{(z/z_0)-1}) = {t \over t_{\epsilon,3}}~,
\ee
where 
\be
t_{\epsilon,3} = {27 d^* \over 8 \sigma^3 \sqrt{z_0}}~ \epsilon^{1 \over 2}~.
\ee
For large times, we get
\be
z_3(t) \approx z_0 \left({t \over t_{\epsilon,3}}\right)^2~.
\ee

For $m=4$, $z_4(t)$ is the solution to
\be
-\ln (z/z_0) + (z/z_0) - 1 = {t \over t_{\epsilon,4}}~,
\ee
where 
\be
t_{\epsilon,4} = {81 d^* \over 16 \sigma^4}~ \epsilon~.
\ee
For large times, we get
\be
z_4(t) \approx z_0 {t \over t_{\epsilon,4}}~.
\ee

For $m=5$, $z_5(t)$ is the solution to
\be
\left((z/z_0)-1\right)^{3/2} - 3 \left((z/z_0)-1\right)^{1/2} + 3 \tan^{-1} \left[ \left((z/z_0)-1\right)^{1/2}\right] 
= {t \over t_{\epsilon,5}}~,
\ee
where 
\be
t_{\epsilon,5} = {243 d^* \over 32 \sigma^5}~ \epsilon^{3/2}~.
\ee
For large times, we get
\be
z_5(t) \approx z_0 \left({t \over t_{\epsilon,5}}\right)^{2/3}~.
\ee

More generally, at large times
\be
z_m(t) \approx z_0 \left({t \over t_{\epsilon,m}}\right)^{2 \over m-2}~,
\ee
\be
a_m(t) \approx a_0 \left({t \over t_{\epsilon,m}}\right)^{1 \over m-2}~,
\ee
where 
\be
t_{\epsilon,m} \propto \epsilon^{m-2 \over 2} \sim {1 \over b(m,\epsilon)}~.  \label{bvbaa}
\ee

From these solutions, it is apparent that the dynamics $z(t) \equiv [a(t)]^2$ is controlled
by the characteristic time $t_{\epsilon,m}$ defined in (\ref{bvbaa}). Note that the inverse 
dependence of $t_{\epsilon,m}$ on $b(m,\epsilon)$ is obvious from the expression 
(\ref{nvbnms}). As the cut-off $\epsilon \to 0$, $t_{\epsilon,m} \to 0$ and the global rupture
occurs in vanishing time. The physical explanation of
this phenomenon is as follows. For $m \geq 2$, the driving force $\sigma^m$ of the damage law 
(\ref{two}) is so strong close to and at the crack tip, that it takes 
effectively zero time for a point to be
brought to the damage threshold. To see this, let us truncate the integral in (\ref{jbbaaa})
such that the upper bound is changed from $t$ to $t-\eta$. The divergence of the
integral at the crack tip means that the contribution to the cumulative
damage occurring in the time interval from $t-\eta$ to $t$ is
larger (actually infinitely larger) than the contribution from time $0$ to time $t-\eta$.
This means that the progressive damage leading to the acceleration of
$a(t)$ for $m<2$ is replaced by an infinite velocity as soon as we start from a
finite crack and do not introduce the finite cut-off length.

This clarifies the physical meaning of the cut-off $\epsilon$ defined in (\ref{bvabbhsk}). A non-zero
$\epsilon$ means that the integral over $\zeta$ in (\ref{bvbbhsk}) does not go all the way up to $z$.
Translated in terms of physical distances, it means that the integral in $\zeta$ in the l.h.s. of (\ref{cnnkfdaz})
also does not go all the way up to $z$. Physically, this means that the damage on a given point
ahead of the crack tip reaches the critical value
$d^*$ before the crack tip reaches that point. The value $d^*$ is no more the rupture threshold but 
a saturation value. 
The crack tip dynamics is now determined by the condition that the damage at any given point $y$ reaches 
this saturation value $d^*$
when the crack tip is at a fixed distance $\propto \epsilon$ from $y$. This condition embodies the 
existence of a microscopic length scale $\propto \epsilon$ such that the damage is no more defined as
smaller scales. 

The main result of our analysis is that the characteristic time scale $t_{\epsilon,m}$ of the crack
dynamics is controlled by the microscopic length scale. The theory has thus fundamentally no continuous limit.
It is one of several interesting and important examples in physics where the macroscopic physics is
completely controlled by the microscopic physics (the ultra-violet cut-off). This situation is found
in many physical problems, for instance in
correlation functions in two-dimensional systems \cite{Patashinski}, 
in non-linear diffusion \cite{Goldenfeld} as well as in 
quantum electrodynamics \cite{bookquantum}. Note however the difference between the last two examples
and the former ones: in our rupture problem as well as in the case of correlation functions in 
2D systems, the ultraviolet cut-off appears
naturally, as an atomic distance, while in the last two case, there is no meaningful natural cut-off,
hence necessity to ``cover-up'' divergencies by the ``renormalization'' procedure \cite{bookquantum}.

\subsection{Regularization by stress saturation at a microscopic scale}

The previous regularization scheme invokes a saturation of the damage at a microscopic length $\propto \epsilon$.
Alternatively, the saturation can occur on the stress field, whose mathematical divergence is bound to 
be rounded off at atomic scales. This provides another regularization scheme.
To implement it, we use the continuous
expression (\ref{three}) for all distances from the crack tip down to a regularization
length $\ell$ such that, for distances from the crack tip from $0$ to $\ell$, the stress is
constant equal to $\sigma(\ell)$ given by (\ref{three}) with $y=a(t)+\ell$. This regularization 
is standard in the theory of damage and of plasticity. The idea is that a sufficiently large damage
exerts a feedback on the stress field which then departs from its damage-free continuous
expression (\ref{three}).
This extension to Rabotnov's treatment provides a natural
way for constructing a self-consistent theory of damage: not only does rupture occur by the 
cumulative effect of damage, damage has also the effect of smoothing out the mathematical
singularity at the crack tip. 
The cut off $\ell$ has the physical meaning of a so-called
process zone or damage zone and its introduction is fully consistent with
the dynamical damage law (\ref{two}). However as we have seen above, for a damage law with an 
exponent $0 < m < 2$, the resulting dynamics becomes insensitive to the existence
of a microscopic length scale in the limit where it is small. In this sense, the regime 
$0 < m < 2$ is more universal and has a continuous limit.

We propose two models that implement these ideas.

\subsubsection{Saturation of the stress at a fixed distance to the crack tip}

The regularization scheme used here is such that
the stress is assumed to saturate at a fixed distance $\ell$ from the crack tip.
Thus, for times $\tau$ up to $t_{\ell}(t)$, the damage at a fixed point that will be reached by the
crack tip at time $t$ is growing under the influence of the stress field created by the crack.
From time $t_{\ell}(t)$ up to time $t$, the damage is increasing linearly with time, since
the stress is assumed constant and equal to the value it reaches at time $t_{\ell}(t)$.
The saturation time $t_{\ell}(t)$ is determined by the equation 
\be
a(t) - a(t_{\ell}) = \ell~.  \label{nvnbvkxk}
\ee
In this version of the regularized theory, expression (\ref{jbbaaa}) is changed into
\be
\int_0^{t_{\ell}} d\tau~ \left({2 \sigma_0 \over 3}\right)^{m}
 {[a(t)]^m \over \left([a(t)]^2 - [a(\tau)]^2\right)^{m \over 2}} 
 +  \left({2 \sigma_0 \over 3}\right)^{m} 
 {[a(t)]^m \over \left[(2 a(t) -\ell) \ell\right]^{m \over 2}} ~(t - t_{\ell})
 = d^*~,   \label{jbbaaaaaaa}
\ee
where $t_{\ell}$ is the time at which the crack tip is at the distance $\ell$ from the
position it will have at time $t$ (see (\ref{nvnbvkxk})).

Note that we now have two equations for two unknown $a(t)$ and $a(t_{\ell})$. The second
term in the l.h.s. of (\ref{jbbaaaaaaa}) expresses the linear increase in damage from 
$t_{\ell}$ to $t$ under the saturated stress. 
In this model, $\ell$ is fixed and $t_{\ell}$ ajusts itself. The value of the saturated stress
is not a constant but increases as the crack gets larger and larger, since it corresponds to the value
at a fixed distance $\ell$ from the tip of a growing crack.

With the change of variables (\ref{nfvnkzka}) and
changing the variable of integration from $\tau$ to $\zeta$ gives
\be
\int_{z_0}^{(\sqrt{z}-\ell)^2} d\zeta~ {(d\tau/d\zeta) \over \left[z-\zeta\right]^{m \over 2}} 
+ {t - t_{\ell} \over \left[(2 \sqrt{z} -\ell) \ell\right]^{m \over 2}}
= \left({3 \over 2 \sigma_0}\right)^m {d^* \over z^{m \over 2}}~~~{\rm with}~~ \sqrt{z}-a(t_{\ell})=\ell~.  
\label{cnnkfafdfdaz}
\ee
Note that $t - t_{\ell}$ can also be written 
\be
\tau(z) -\tau((\sqrt{z}-\ell)^2) \approx 2 \ell \sqrt{z} ~{d\tau \over dz}|_z + {\cal O}\left(\ell^2\right)~.
\label{ngnksk}
\ee

The integral in the l.h.s. of (\ref{cnnkfafdfdaz}) is analyzed similarly to the previous case (\ref{cnnkfdaz}).
We multiply
the integral by $1/(y-z)^{1-{m \over 2}}$ and integrate over $z$ from $(\sqrt{z_0}+\ell)^2$ to $y$:
\be
\int_{(\sqrt{z_0}+\ell)^2}^{y} dz \left(\int_{z_0}^{(\sqrt{z}-\ell)^2} d\zeta~ {(d\tau/d\zeta) \over 
\left(z-\zeta\right)^{m \over 2} (y-z)^{1-{m \over 2}} }\right) =
\int_{z_0}^{(\sqrt{y}-\ell)^2} d\zeta {d\tau \over d\zeta} \left[
\int_{(\sqrt{\zeta}+\ell)^2}^{y} {dz  \over \left(z-\zeta\right)^{m \over 2} (y-z)^{1-{m \over 2}}}\right]~,  
\label{cbvvaa}
\ee
where we have used the equality of the triangle 
$\int_{(\sqrt{z_0}+\ell)^2}^{y} dz  \int_{z_0}^{(\sqrt{z}-\ell)^2} d\zeta = 
\int_{z_0}^{(\sqrt{y}-\ell)^2} d\zeta \int_{(\sqrt{\zeta}+\ell)^2}^{y} dz$.

The integral in the bracket in the r.h.s. is the same as in (\ref{bvabbhsk}), which defines the function 
$b(m,\epsilon)$  with
\be
\epsilon(\zeta) = {(\sqrt{\zeta}+\ell)^2 - \zeta \over y-\zeta} \approx {2 \ell \sqrt{\zeta} \over y - \zeta}~.
\ee
Note that $\epsilon(\zeta)$ is now a function of $\zeta$. 

Using (\ref{ngnksk}) and (\ref{cbvvaa}), expression (\ref{cnnkfafdfdaz}) gives
\be
\int_{z_0}^{(\sqrt{y}-\ell)^2} d\zeta ~{d\tau \over d\zeta} ~b(m, \epsilon(\zeta))
+ \int_{(\sqrt{z_0}+\ell)^2}^{y} dz {2 \ell \sqrt{z} \over (y-z)^{1-{m \over 2}} 
\left[(2 \sqrt{z} -\ell) \ell\right]^{m \over 2}} ~{d\tau \over dz}
= \left({3 \sqrt{d^*} \over 2 \sigma_0}\right)^m \int_{(\sqrt{z_0}+\ell)^2}^{y}   
{d\zeta \over \zeta^{m \over 2} (y-\zeta)^{1-{m \over 2}}}~.   \label{nvgjes}
\ee
Since 
\be
b(m, \epsilon(\zeta)) \propto 1/\epsilon^{{m \over 2}-1} \propto \left({2 \ell \sqrt{\zeta} 
\over y - \zeta}\right)^{1-{m \over 2}}~,
\ee
we see that the first integral of the l.h.s. of (\ref{nvgjes}) is negligible compared to the second
integral of the l.h.s. of (\ref{nvgjes}) in the limit of large cracks, i.e. large $z$.
Neglecting $\ell$
compared to $\sqrt{z}$ in the denominator of the integrant 
of the second integral of (\ref{nvgjes}) and equating this second integral to the r.h.s. gives the 
following equation
 \be
 {d\tau \over dz} = (2\ell)^{{m \over 2}-1}~ \left({3 \sqrt{d^*} \over 2 \sigma_0}\right)^m ~
 {1 \over z^{{m \over 4}+{1 \over 2}}}~.   \label{hgaknav}
 \ee
 For $m > 2$, ${m \over 4}+{1 \over 2} > 1$ and the solution of (\ref{hgaknav}) is
 \be
 a(t) \propto {\ell \over (t_c-t)^{2 \over m-2}}~,  \label{bgakdsna}
 \ee
 where $t_c$ is determined from the initial size of the crack.
 The finite-time singularity results from the ever-increasing stress field at the fixed
 distance $\ell$ from the crack tip. This solution (\ref{bgakdsna}) is 
 qualitatively different from the solution (\ref{hgkzz}) found in the regime $m < 2$
 as (\ref{bgakdsna}) depends in a fundamental way upon the existence of the regularization scale 
 $\ell$.

\subsubsection{Saturation by fixing an absolute maximum stress}

An alternative prescription for the regularization is that the stress saturates at a constant value
$\sigma_{\rm max}$. This is in constrast with the previous regularization scheme where the
stress saturates at a value reached at a constant distance, this value thus increasing with the crack length.
Expression (\ref{jbbaaaaaaa}) is then changed into
\be
\int_0^{t_{\ell}} d\tau~ \left({2 \sigma_0 \over 3}\right)^{m}
 {[a(t)]^m \over \left([a(t)]^2 - [a(\tau)]^2\right)^{m \over 2}} 
 +  [\sigma_{\rm max}]^m~(t - t_{\ell}) = d^*~,   \label{jbbaaadaaaa}
\ee
with
\be
{2 \sigma_0 \over 3}~
 {a(t) \over \left( [a(t)]^2 - [a(t_{\ell})]^2 \right)^{1 \over 2}} = \sigma_{\rm max}~,  \label{fjakka}
 \ee
which is the condition that the stress saturates. It gives
\be
[a(t_{\ell})]^2 = [a(t)]^2 ~(1-A^2)~,
\ee
where
\be
A = {2 \sigma_0 \over 3 \sigma_{\rm max}}~.
\ee
The equation (\ref{jbbaaadaaaa}) governing the dynamics of the crack tip can thus be
written
\be
{2 \sigma_0 \over 3} \int_{z_0}^{(1-A^2)z} d\zeta~ {d\tau/d\zeta \over
(z-\zeta)^{m/2}} + [\sigma_{\rm max}]^m~A^2~z~{d\tau \over dz} = d^*~,   \label{jbzlmaaaa}
\ee
where we have used the expansion
\be
\tau(z) -\tau((1-A^2)z) \approx A^2 ~z ~{d\tau \over dz}|_z + {\cal O}\left(A^4\right)~,
\label{ngaanksk}
\ee
valid in the interesting regime $\sigma_{\rm max} >> \sigma_0$ giving $A << 1$.

The second term in the l.h.s. of (\ref{jbzlmaaaa}) dominates the first integral for large $z$, as 
can be checked a posteriori. For large crack sizes, the expression (\ref{ngaanksk}) can
thus be simplified into 
\be
[\sigma_{\rm max}]^m~A^2~z~{d\tau \over dz} =  d^*~,   \label{jbaazlmaaaa}
\ee
whose solution is
\be
a(t) = a_0 ~e^{t/t_0}~,
\ee
where
\be
t_0 = {2 d^* \over A^2 [\sigma_{\rm max}]^m} = {9 d^* \over 2 \sigma_0^2~[\sigma_{\rm max}]^{m-2}} ~.
\ee

\section{Beyond the mean field version by functional renormalization}

Let us restrict our discussion to the case $m=1$ for which we have the complete analytical solution for
the crack dynamics. The solution (\ref{njjakak}) with its asymptotic behavior 
(\ref{pow}) is not physically reasonable, as
the crack reaches an infinite length in a finite time. The $(t_c-t)^{-1}$ singularity
has been found to appear as the consequence of a geometric nonlinearity on an otherwise  
linearized mechanical problem. In reality, nonlinearity,
viscosity, feedback, spatial heterogeneity of material properties and of cracking
should modify the singularity. In addition, the main simplification in the previous approach
is to neglect the impact of damage on the elastic coefficients of the material, thus leading
to a stress field created by the crack which is identical to the field that the same {\it static} crack 
would generate in an undamaged material. 
Our hypothesis is that 
such modification can be deduced by a smooth or regular deformation of the
solution previously obtained. 

In this goal, we propose to apply the Yukalov-Gluzman functional renormalization method \cite{YG}
to the series expansion of the solution (\ref{njjakak}) to obtain the renormalized law
that accounts for these effects in a generic sense. Let us first consider the
asymptotic power law singularity (\ref{pow})
\be
a(t) = {2 a_0 \over \pi} ~(1-x)^{-1}~,~~~~~~{\rm where}~~ x \equiv t/t_c~.    \label{fnakaa}
\ee
The powers $x^n$ in the expansion
\be
a(t) = {2 a_0 \over \pi} ~\left( 1+x + x^2 + x^3 + ... \right)   \label{nvkkcvm}
\ee
may be considered as hidden free parameters. Indeed, let us
multiply the expansion by $x^s$. We then have a trial expansion for the solution.
For $s=0$, we return to the regular expansion. Such multiplication can be applied
repeatedly, for instance using the functional renormalization method \cite{YG}.
The idea behind the introduction of the multiplicative (control) function such as
the power $s$ in  $x^s$
is to deform smoothly the initial functional space of the expression $a(t)$ taken as
an approximation to be improved. The condition for the improvement is 
to obtain a faster and better controlled convergence in the space of the
modified functions upon addition of successive terms $x^n$ in the expansion. 
By this procedure, the dominant poles are eliminated or weakened 
as a result of a sequential reduction of stress level at each step of
the resummation procedure. This corresponds to utilizing the information
from the initial series pertaining to the times preceding the critical time $t_c$, where 
the level of damage is lower. Thus, the renormalization procedure is performing a mapping
from the dynamics at early time far from the critical point to later times closer to the
critical time. The stabilization stems from the fact that the
information contained in the initial series related to times  close to $t_c$ is minimized on
the basis that it has an overly destabilizing effect in the description and 
should be weighted less than the information at earlier times.

At each step of the functional renormalization corresponding to the addition of a new term, 
we select the renormalized function according to
the principle of minimum ``local'' multiplier, i.e., maximum stability on each sub-step of
the renormalization procedure. Since these multipliers are proportional to
the derivative $da/dt$ \cite{YG}, the principle of minimal multiplier implies a selection of the
real-time trajectory of the crack with minimal rate of damage (minimal stresses). In other words, this procedure
amounts to improve the theory by allowing the crack to organize and develop so as
to choose the most favorable path or dynamics. It can be shown \cite{YG} that, 
at each step, the choice of a formally infinite exponent $s$ corresponds to the minimal multiplier
at arbitrary time.

The functional form of a super-exponential solution is selected by this procedure: starting
from an expansion $1+ a_1 x + a_2 x^2 + a_3 x^3 +...+a_k x^k$, the renormalized expression is as follows.
With the use of the notation
\begin{equation}
\label{39}b_0=a_{0\ },\quad b_k=\frac{a_k}{a_{k-1}}\ ,\ \ k=1,2,...,
\end{equation}
we obtain the {\it bootstrap self-similar approximant} up to order $k$
\begin{equation}
\label{40}
F_k(x)=b_0\exp (b_1x\exp (b_2x\exp (...b_{k-1}x\exp (b_kx)))...)~,
\end{equation}
introduced by Yukalov and Gluzman \cite{YG}. 

Let us now apply this result to the case (\ref{nvkkcvm}) where
all the coefficients $a_n$ are equal to $1$.
The corresponding renormalized approximant replacing the initial input $1/(1-x)$ of the expansion reads
\begin{equation}
\label{f18aaaa} F(x)= \exp \left( x \exp \left(  x ...\exp \left(  x \right) \right)...\right)~ .
\end{equation}
This embedded exponential series converges to a well-defined function. To determine it, we note that
$F(x)$ obeys the recursion relation
\be
F_{k+1}(x) = \exp \left( x F_k(x) \right)~.
\ee
The fixed point to which these series of approximants converge is thus solution of
\be
F = \exp \left( x F \right)~,  \label{fjabfa}
\ee
noting $F= \pi a / 2 a_0$.
The limit $F_{\infty} (x)$ exists for $-{\rm e} \leq x \leq {\rm e}$.

The fixed point $F(x)$ can be shown \cite{benderorszag} to be the solution of the
equation
\be
{d F \over dx} = {F^2 \over 1 -x F}~.
\ee
Searching for a solution in the form of a Taylor series
\be
F(x) = \sum_{n=0}^{\infty} y_n x^n~,~~~~y_0 = 1~,  \label{nvbkjx}
\ee
we get
\be
y_n = {(n+1)^{n-1} \over n!}~.
\ee
Since $n! \approx n^n e^{-n}$, $y_n \approx e^n$ for large $n$ and the generic term
$y_n x^n$ in the series (\ref{nvbkjx}) is proportional to $(ex)^n$. This shows that 
the radius of convergence of the series (\ref{nvbkjx}) is $1/e$.

$F(x)$ has a singularity when $x$ approaches $1/e$ from below, whose shape is obtained by expansions of 
 expression (\ref{fjabfa}):
\be
F(x) =_{x \to 1/e}~ e \left( 1 - \sqrt{2}~e^{3/2} \sqrt{1/e - x}\right)~.
\label{fnlala}
\ee

Thus, the self-similar functional renormalization has transformed a pole (divergence of $a$) at $t=t_c$
into a square root singularity (finite $a$) at a smaller $t=t_c/e$. In this renormalized theory,
the crack accelerates up to the time $t_c/e$ as which time its velocity diverges, while the crack
is still finite. This announces the global breakdown. It is interesting that the exponent $1/2$ is close
to the value found for acoustic emissions in experiments \cite{Anifrani,Ciliberto,Anifrani2,criticalrecent}.

We can offer the following physical intuition for this transformation from the solution (\ref{pow}) with 
$\beta=1$ to to $\beta=-1/2$. As the material becomes more and more damaged, the ulterior
functional dependence of damage as a function of applied stress is modified. Actually, the
series of functional renormalization amounts to effectively evolve or renormalize the
damage law (\ref{two}) into a succession of effective laws captured by the sequence of 
approximants, each approximant order corresponding to an increase in the overall damage 
of the material. Here, we have a mapping between a measure of evolution via the 
cumulative damage, i.e., a measure of passed time, and the order of the approximants and thus
the distance to the fixed point in the functional space.

Consider now the general case $a(t) \sim (1-x)^{-\beta}$.
Expanding in power series, we get
\be
(1-x)^{-\beta} = \sum_{n=0}^{\infty}  a_n x^n~, 
\ee
where
\be
a_n = {(n+\beta-1)! \over n!(\beta-1)!}~.
\ee
The Yukalov-Gluzman renormalization scheme gives the superexponential (\ref{40})
with coefficients $b_n$ given by
\be
b_n = (n+\beta-1)/n ~.
\ee   
Since $b_n \to 1$ for large $n$ for any $\beta$, the fixed point of the approximants is
controlled by the same finite square-root singularity of the type (\ref{fnlala}). 
Thus, the functional renormalization maps all finite-time singularities with different 
exponent $\beta$ on the same universal law $a(t) = a(t_c) - C \sqrt{t_c -t}$, where $C$ is
a constant depending in particular on $\beta$.

\section{Concluding remarks}

Two main regimes have been found for the growth of a crack 
in a medium obeying the damage law $d(d)/dt = \sigma^m$
(equation (\ref{two})), where $\sigma$ is the local stress.
For $0 < m < 2$, a pre-existing crack grows to infinity in finite time and the divergence occurs
as a power law finite-time singularity. For $m \geq 2$, the solution exists for all times but the
characteristic time scale of the crack growth is an increasing function of a microscopic length scale, 
which is essential for regularizing the otherwise ill-defined problem. This microscopic length
scale embodies the physical mechanism(s) by which the mathematical stress singularity at the crack
tip of a perfectly sharp crack is rounded-off. We have examined two main scenarios, a 
damage-limited rupture and a stress-limited rupture.   

The remarkable behavior of this simple model results from the form of the irreversible damage law, 
in particular from the fact that any non-vanishing stress increases the damage. Damage at any
point is thus a kind of
ever increasing counter of the history of the stress on that point. This feature prevents the 
existence of stationary solutions of cracks propagating at constant velocities. In contrast, we only
obtain ``run-aways.'' 

Stationary solutions can be obtained in
simple generalizations of the damage law (\ref{two}), for instance with a stress threshold 
below which no damage occurs or with a healing or work-hardening term allowing recovering of the material
and decrease of the damage when the stress is low. Such situations have been investigated in
discrete two-dimensional models \cite{mechaanalog}.

\pagebreak

\begin{figure}
\begin{center}  
\epsfig{file=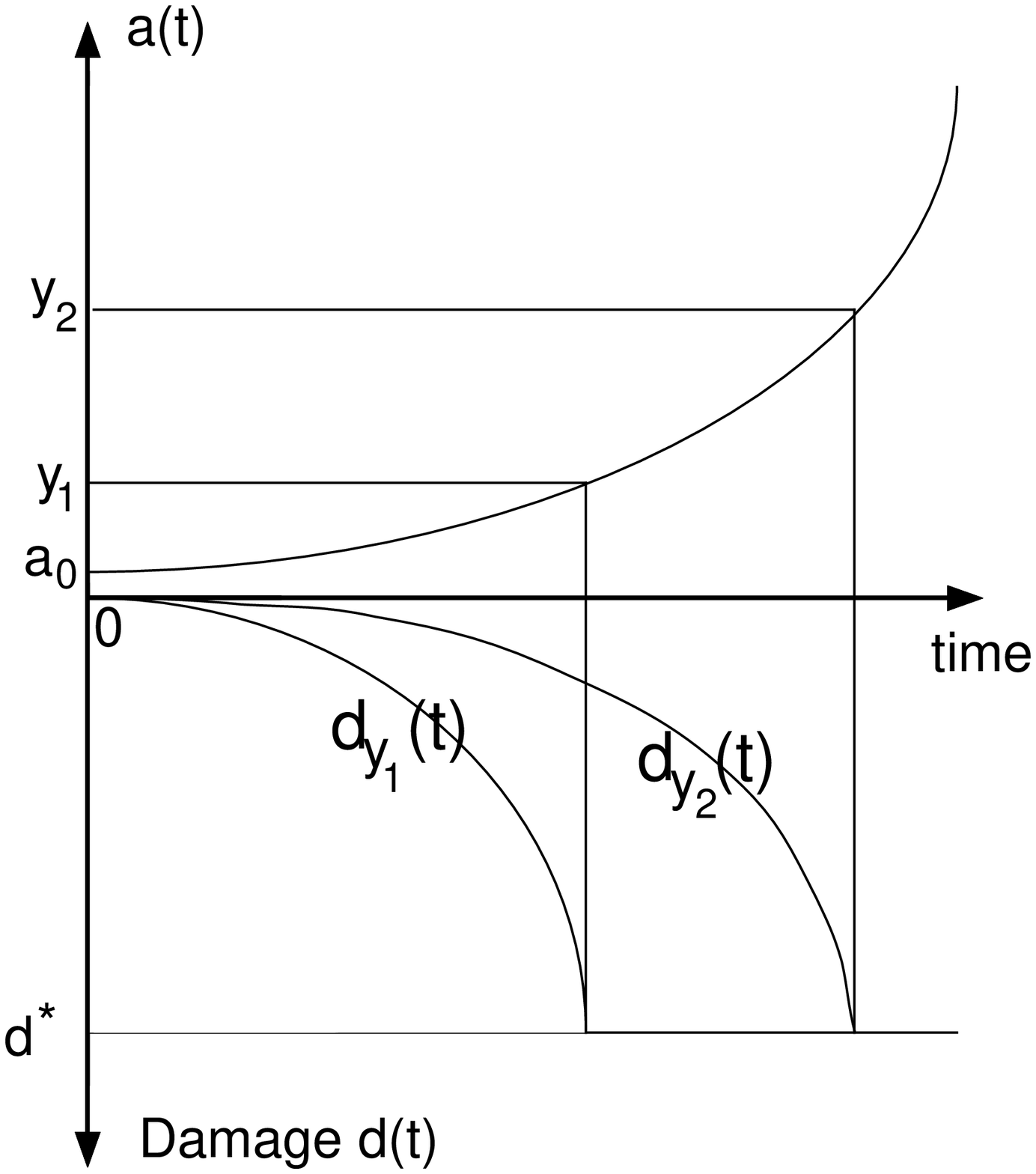,height=17cm,width=14cm}
\caption{\label{fig1} Illustration of the law
governing the growth of the crack: the
dynamics of its length $a(t)$ is obtained from the self-consistent condition that the time it
takes from a point at $y$, at the distance $y-a(\tau)$ from the crack tip at time
$\tau$, for its damage to reach the rupture threshold $d^*$ is exactly equal to
the time taken for the crack to grow from size $a(\tau)$ to the size $a(t)=y$ so
that its tip reaches the point $y$ exactly when it ruptures. }
\end{center}
\end{figure}

\end{document}